\documentclass[twocolumn,pra]{revtex4-1}
\usepackage[T1]{fontenc}
\usepackage[utf8]{inputenc}
\usepackage{amsmath,amssymb}
\usepackage{color}
\usepackage{graphicx}
\usepackage{bm}
\begin{document}

\title{Edge states of moir\'e structures in graphite}
\author{E. Su\'{a}rez Morell}
\email{eric.suarez@usm.cl}
\affiliation{Departamento de F\'{i}sica, Universidad T\'{e}cnica
Federico Santa Mar\'{i}a, Casilla 110-V, Valpara\'{i}so, Chile}
\author{P. Vargas}
\affiliation{Departamento de F\'{i}sica, Universidad T\'{e}cnica
Federico Santa Mar\'{i}a, Casilla 110-V, Valpara\'{i}so, Chile}
\author{P. H\"aberle}
\affiliation{Departamento de F\'{i}sica, Universidad T\'{e}cnica
Federico Santa Mar\'{i}a, Casilla 110-V, Valpara\'{i}so, Chile}
\author{S. A. Hevia}
\affiliation{Instituto de F\'{i}sica, Pontificia Universidad Cat\'olica de Chile,
Avda. Vicu\~na Mackenna 4860, Macul, Santiago, Chile}
\author{ Leonor Chico}
\affiliation{Departamento de Teor\'{\i}a y Simulaci\'on de Materiales, Instituto de Ciencia de Materiales de Madrid (ICMM),
Consejo Superior de Investigaciones Cient\'{\i}ficas (CSIC),
C/ Sor Juana In\'es de la Cruz 3,
28049 Madrid, Spain}

\date{\today}
\pacs{}

\begin{abstract}
We address the origin of bead-like edge states observed by scanning tunneling microscopy (STM) in  moir\'e patterns of graphite.
Low-bias STM measurements indicate these edge states are centered around AB stacking sites, contrarily to the common assumption of them being at AA sites.  This shift of the beads intensity with respect to the bulk moir\'e pattern has been corroborated
by a
tight-binding
calculation of the edge states in bilayer flakes. Our results are valid not only for graphite but also for few-layer graphene, where these states have also been recently observed.

\end{abstract}

\maketitle

\section{Introduction}
%


Structural moir\'e patterns can be found in materials in which two superimposed lattices,  with different lattice constant and/or a relative rotation between them are present.
With the advent of two-dimensional solids and the possibility of stacking them in diverse ways \cite{Geim_2013}, the influence of these moir\'e superstructures in the electronic structure  has to be taken into consideration for a complete description of these materials.
Indeed, very recent experiments of graphene over hexagonal boron nitride show an unexpected insulating behavior 
 that might ultimately be related to the moir\'e superlattice \cite{Amet_2013,Jarillo_2012,Woods_2014}.

The physical properties of structures composed of few-layer graphene \cite{Novo_2004} have been shown to display a
dependence on the number of layers.
Specifically, when two stacked graphene layers are rotated with respect to each other, a distinctive moir\'e pattern, linked to the electronic density,  develops in the $\pi$ electronic states of the bilayer, being sharper for small rotation angles. In fact, an intriguing dependence with the relative rotation angle (RRA) between layers has been observed in
STM measurements. For large  RRA ($\thicksim 20^{\rm o}$),
  the system behaves as if the two layers were uncoupled and a linear energy dispersion in momentum is retained, like in monolayer graphene. For RRA below $20^{\rm o}$
   the velocity of the charge carriers is reduced  
 and the STM moir\'e pattern becomes brighter. Scanning tunneling spectroscopy (STS) measurements have shown an angular dependence of the Van Hove singularities consistent with the development of zero-velocity flat bands.
These peculiar features have attracted the attention of the graphene community,
and many efforts have been dedicated to study and explain the properties of twisted bilayer graphene and the RRA dependence of its physical properties  \cite{Lopes_2007,Shallcross_2008, Shallcross_2010,Trambly_2010,Morell_2010,Mele_2010,Bistritzer2011,Lopes_2012,SanJose_2012,Trambly_2012a,Li_2010,
Luican_2011,Trambly_2012b,Xian_2011,Morell_2013,Morell_2014a,Morell_2014b}.

Besides bilayer graphene, highly oriented pyrolytic graphite (HOPG)
is another example of material in which this type of moir\'e
structures has been observed
 \cite{Kuwabara_1990,Rong_1993,Xhie_1993,Rong_1994,Ouseph_1996,Bernhardt_1998,Osing_1998,Ouseph_2011,Pong_2005}.
HOPG comprises a large number of micron-size crystallites having random in-plane orientations, and in each crystallite, the graphene layers are stacked in an ABA sequence. The  hexagonal unit cell has  two non-equivalent sites;  the $\alpha$ site, where a carbon atom is directly placed above an atom of the underlying layer, and the $\beta$ site, for which the carbon atom lies on a hollow site at the center of an hexagon in the layer below. The difference in the local density of states (LDOS) at both sites explains why STM images of HOPG display a three-fold symmetry instead of six-fold \cite{Tomanek_1988}. The $\beta$ sites, with larger LDOS, are the bright spots in STM images, while the $\alpha$ sites are dark due to a reduced LDOS.
Nevertheless, under some experimental conditions, the six bright spots can been observed \cite{Wang_2006,Cisternas_2008}.

Sometimes,  a top layer graphene sheet on HOPG is misoriented with respect to the underlying solid, giving rise to moir\'e patterns, i.e., forming a superlattice with periodicity in the nanometer range.
For several years the origin of these moir\'e patterns remained a controversial topic, but currently,  it is widely accepted that these superstructures in graphitic materials  are directly linked to the electronic structure of the rotated
layers. 
In fact, an interesting debate arose with respect to which site in the unit cell was responsible for the bright areas in STM measurements.
As bright STM spots measured in Bernal (ABA) HOPG were related to AB sites, some authors also associated the bright areas in moir\'e patterns to AB stacking. However, density functional theory (DFT) and tight binding (TB) simulations of the electronic structure revealed that the AA site is indeed responsible for the bright regions in STM images \cite{Xhie_1993,Rong_1993,Campanera_2007}.

Another 
intriguing aspect of these topographic images are the so-called bead-like features, related to edge states in carbon-based moir\'e structures. A rotated graphene flake over 
HOPG presents brighter spots at the edges than in the moir\'e pattern observed inside
the flake. By using STM, Berger {\it et al.}   \cite{Berger_2010}
have shown that these edge states do not necessarily follow the symmetry of the bulk moir\'e pattern if the topographic measurements are performed with a bias voltage of 500 mV.  Under these tunneling conditions, a "disordered" decoration on the edges of the top layer is observed.
Despite the fact that these bead-like edge states have been detected several times in the last few years \cite{Xhie_1993,Sun_2003,Pong_2007,Varchon_2008,Berger_2010,Zhang_2013}, an explanation of their origin is still absent. 

In this work we
provide a consistent explanation
of the bead-like features observed at low bias in STM images.
The analysis of STM images of these low-energy features indicates that the bright spots at the edges are not located at  AA sites, as it is the case in
bulk moir\'e patterns, but instead they are located at AB edge sites. We explain this result by simulating the LDOS of a rotated flake over graphene and analyzing the energy distribution of  their edge states.
The LDOS of the edges have maxima at different energies, depending on the edge stacking: AA or AB.  The edge states with energies closer to the Fermi level ($E_F$) turn out to be brighter in AB edge sites, in agreement with our STM observations.



\section{Experimental method}

\begin{figure}[htbp]
\includegraphics[width=\columnwidth,clip]{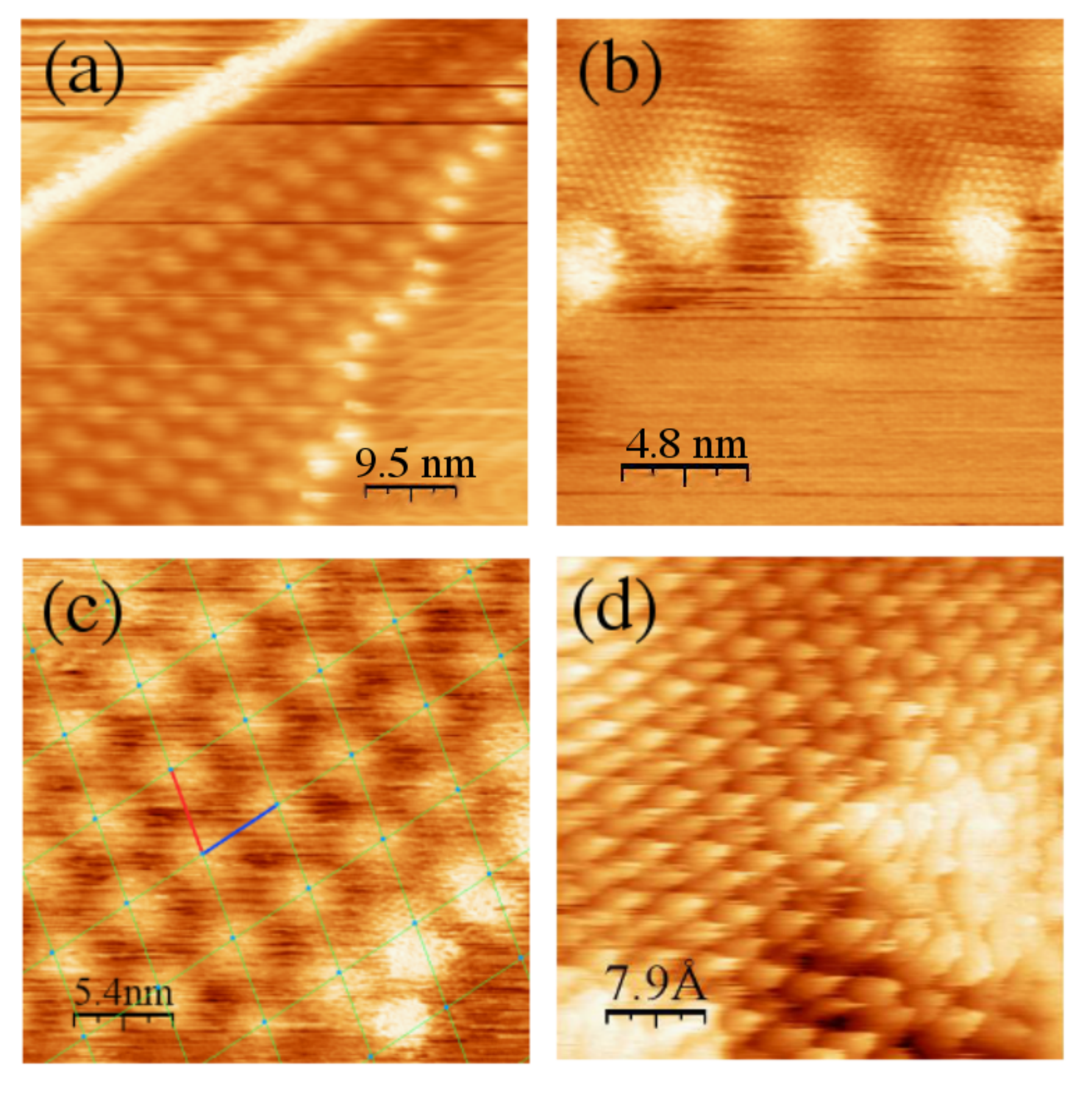}
\caption{(Color online) Typical STM image of a moir\'e superlattice in HOPG collected with a bias voltage of 100 mV.  Both edge states and bulk moir\'e patterns are clearly observed. (a) 
Wide overview of the surface; (b) detailed section of the edge, where it is clear that the edge states do not follow the symmetry of the superlattice; (c) detailed view of the bulk moir\'e  pattern, used to estimate the size of the supercell; (d) detailed view of the coexistence of the triangular atomic unit cell of graphite and the brighter moir\'e intensity induced by the layer rotation.  
}
\label{Fig1}
\end{figure}

The STM images shown in Fig. \ref{Fig1} were taken from HOPG cleaved in air \cite{HOPGsample}. The sample was swiftly introduced  into an ultra-high vacuum chamber (base pressures below $10^{\rm -10}$ Torr) equipped with a commercial Omicron scanning tunneling microscope. Topographic images were collected at room temperature in a constant current mode, adjusted to 2 nA, and a bias voltage of 100 mV. Nanotec Electronica WSXM software was used for image processing \cite{WSXM_2007}. Fig. \ref{Fig1}(a) shows a region with a distinctive moir\'e structure, in which bead-like edge states are clearly distinguished. In Figs. \ref{Fig1} (b) and (d) it is possible to observe not only the brighter moir\'e pattern spots, but also the fainter triangular lattice image displayed by AB stacked graphite in topographic STM measurements. 
The bright spot in the moir\'e cell corresponds to the AA region; the rest of the cell is in a large percentage composed of AB-like regions \cite{Campanera_2007}. As a result one observes the well-known triangular lattice of AB-stacked graphite between the bright spots corresponding to AA sites. The periodicity of the
moir\'e pattern
was obtained by superimposing a periodic grid over the image in panel (c).  From this graphical procedure, we were able to measure
the length of the supercell, which is $D=4.8$ nm.
The periodicity of the supercell is related to the RRA between layers $\theta$ by the equation $D= a_{0} / 2 \times \sin(\theta/2)$, where $a_{0}=2.46$ \AA\
is the length of the graphene lattice vector. 
For this particular pattern this expression yields a value of $\theta=2.9^{\rm o}$.

In Figs. \ref{Fig1} (b) and (c), one can observe that the brighter edge states do not follow the periodicity of the bulk moir\'e pattern.
This misalignment of the bright beads at the edges of rotated graphene layers arises from the combination of the moir\'e enhancement of the local density of states and the presence of edge states due to the termination of the upper graphene layer. With the purpose of explaining these observations, in the following section we describe some theoretical issues related to this system.


%
\section{Theoretical considerations}
 \label{sec:theory}

In order to explain the STM images obtained from HOPG it is sufficient to consider only two graphene layers. The relative rotation of one layer with respect to the other is enough to reproduce the moir\'e pattern observed in infinite systems.
\subsection{Geometry}
The unit cell of two rotated graphene layers, the so-called twisted bilayer graphene (TBG), is obtained by rotating a Bernal-stacked bilayer.
As a reference we choose the rotation axis to be in a B site, which has one atom in the top layer situated on a hollow site, at the center of a hexagon in the layer underneath \cite{Morell_2011b}, but it can
equivalently be taken to be in an A site, where two atoms are exactly on top of each other \cite{Campanera_2007}. We briefly summarize the notation employed for these structures.
A commensurate unit cell (UC) for TBG is built as follows: a point of the crystal with coordinates $\mathbf{r}=m \mathbf{a}_{1} + n \mathbf{a}_{2}$ (\textit{n},\textit{m} integers) is rotated to an equivalent site $\mathbf{t}_{1}=n \mathbf{a}_{1} + m \mathbf{a}_{2}$. Here
 the graphene lattice vectors are given by  $\mathbf{a}_{1}=\frac{a}{2}(\sqrt{3},-1)$ and  $\mathbf{a}_{2}=\frac{a}{2}(\sqrt{3},1)$, and $a=2.46\,$\AA\  is the graphene lattice constant. The  vectors of the unit cell for TBG can be chosen as $ \mathbf{t}_{1}=n \mathbf{a}_{1} + m \mathbf{a}_{2} $ and $\mathbf{t}_{2}= -m \mathbf{a}_{1} + (n+m) \mathbf{a}_{2}$.
The corresponding twisted bilayers 
are usually labeled by the indices $(n,m)$ \cite{Morell_2010,Trambly_2010}.  The distance between layers is set to $3.35$ \AA.

\begin{figure}[htbp]
\includegraphics[width=\columnwidth,clip]{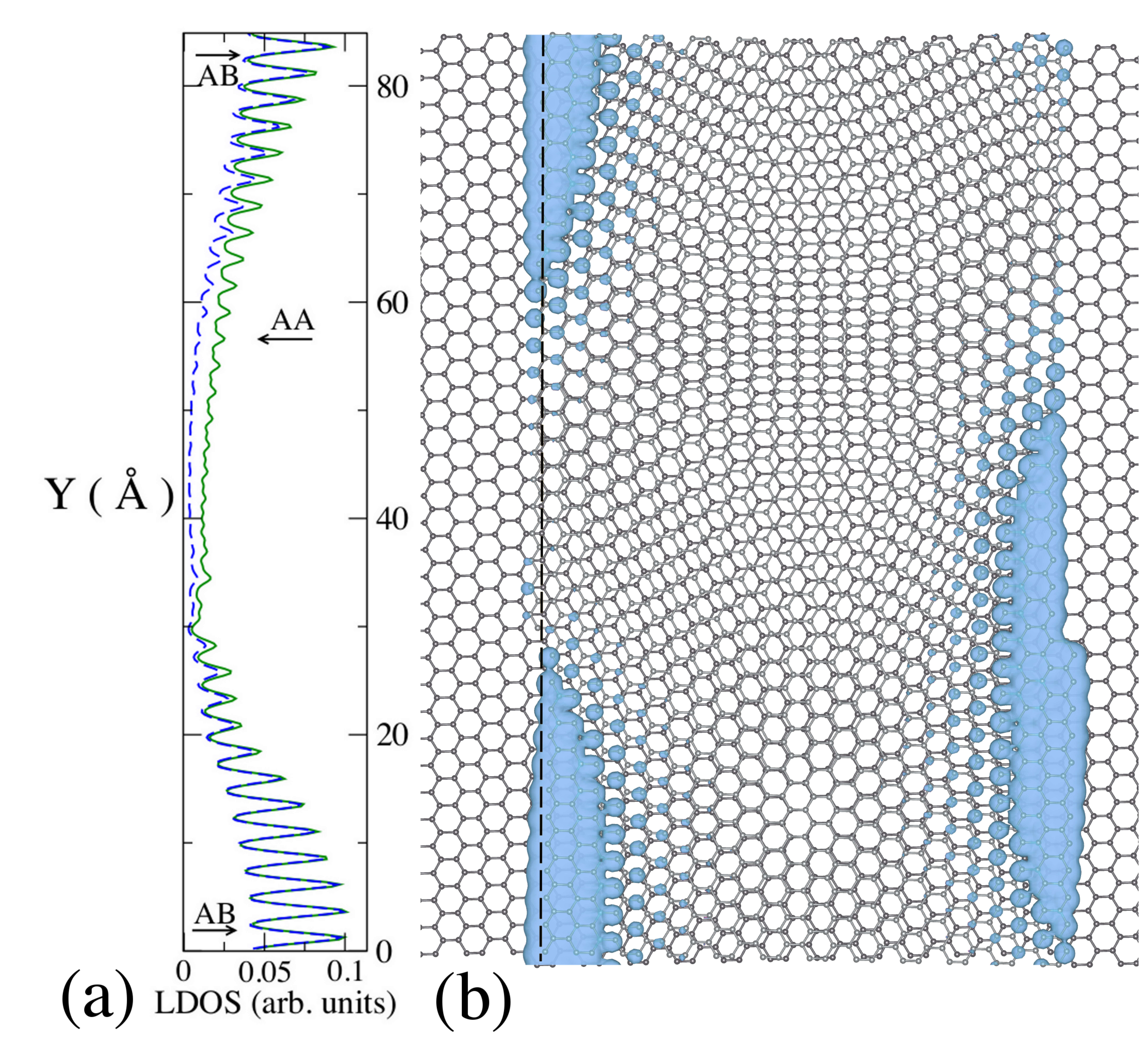}
\caption{(Color online)  STM (LDOS) simulation of edge states. (a) STM line profile along the left edge of the terrace. In dashed blue we indicate the profile for energies $E$ in the range $E_{F}-100$ meV < $E$ < $E_{F}$ and in solid green for $E_{F}-500$ meV < $E$ < $E_{F}.$ (b) Simulation of the STM topographic profile for a tip voltage of 100 mV. The line profile depicted in (a)  was measured along the dashed line close to the edge of  the top flake.  }
\label{Figribbon}
\end{figure}

The repetition of a TBG unit cell with $n=m+1$ along either $ \mathbf{t}_{1}$ or  $ \mathbf{t}_{2}$, yields a chiral ribbon with edges being predominantly armchair. However, since we focus in edge states, which are produced by zigzag edge atoms \cite{Jaskolski_2011}, we choose a unit cell along the direction perpendicular to  $ \mathbf{t}_{1}$. This particular choice doubles the size of the unit cell, yielding a ribbon with a majority of zigzag edge atoms \cite{Morell_2014b}. To perform our calculations we extend the size of the lower ribbon to simulate an infinite graphene. The system behaves then like a graphene layer with a flake on top rotated with respect to the lower layer. The right panel of Fig. \ref{Figribbon} shows the geometry of the system with a (12,11) moir\'e pattern. The graphene layer has been cut for illustrative purposes. In the figure is also shown the LDOS of all states integrated from  $E_{F} -100$ meV to the Fermi energy. The edges are chosen to be minimal, i.e., with a minimum number of atoms, all with coordination number 2. The RRA for this structure in $2.88^{\rm o}$,  very close indeed to the actual value experimentally observed.

\subsection{Tight-Binding Model}
\label{sec:TB}
The electronic properties of these bilayer graphene ribbons are modeled within the tight-binding approximation including only $p_{z}$ orbitals \cite{Morell_2011b}.  Within each layer, we considered a fixed nearest-neighbor intralayer hopping parameter  $\gamma_0=-3.16$ eV. For the interlayer interaction we used a distance-dependent
 hopping \cite{Shallcross_2010,Trambly_2010,Morell_2010}.
  Thus, the Hamiltonian is
given by $H=H_{1}+H_{2}+H_{12} $, where $H_{n}$ ($n=1,2$) correspond  to the Hamiltonian of each layer
and $H_{12}$ describes the interlayer coupling,
$H_{12} =  \sum_{i,j} \gamma_{1}e^{- \beta ( r_{ij}-d)} c^\dagger_{i} c_{j} + H.c.$,
where $\gamma_{1}=-0.39$ eV is the nearest-neighbor interlayer hopping energy scale, $d$ is the interlayer distance, $r_{ij}$ is the distance between atom $i$ on layer $1$ and atom $j$ on layer $2$, and $\beta =3$.
This value of $\beta$ accurately reproduces the dispersion bands calculated within a density functional theory approach \cite{Morell_2010,Shallcross_2008}.
We set a cutoff for the interlayer hopping equal to
 $6 a_{CC}$, with $a_{CC}=1.42$ \AA\ being the nearest-neighbor distance between carbon atoms.


\section{Discussion}
 \label{sec:remarks}

Figs. \ref{Figribbon} (a) and (b) show that the LDOS of states with  energies very close to the Fermi level 
are 
mainly located at AB-stacked edge regions, whereas the AA-stacked sections of the edges do not show any appreciable density of states.
Fig. \ref{Figribbon} (a)
presents a line profile of the LDOS along the left edge of the flake.
Arrows indicate the
position
of
AA and AB sites at the edge. 
Clearly, the LDOS is higher at the AB sites. 
Calculations
were
made integrating the LDOS  from 100 meV below the Fermi level to $E_F$
(dashed line) and for states with energies between $E_F -500$ meV and $E_F$ (solid line). 
The LDOS is higher at the AB sites in both cases, but for the
wider integration range
an increase in the LDOS at the AA sites is observed. This simulation implies that when a topographic STM image of such system is taken with tip voltages around 100 mV, bright regions at the edges should correspond to AB sites, while non-edge areas are brighter in AA regions. For large tip voltages, around 500 mV, the bright areas at the edges will be also the AB sites, but under some experimental conditions the AA sites might also be visible, creating an smearing pattern in some regions \cite{Berger_2010}.

This effect can also be explained in a qualitative manner using the band structure of bilayer graphene nanoribbons with AA and AB stacking \cite{Morell_2014b}, together with the different coupling strengths in AA and AB-stacked edges.

\begin{figure}[htbp]
\includegraphics[width=\columnwidth,clip]{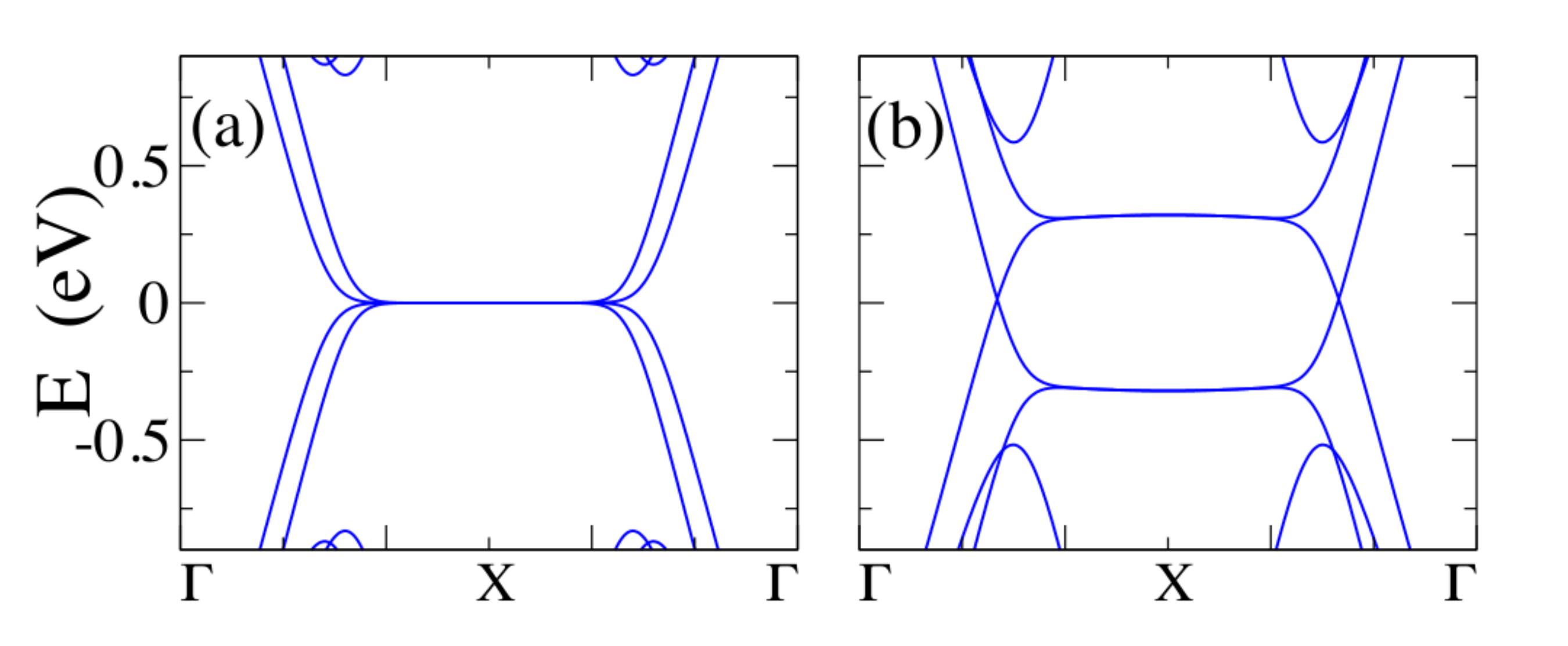}
\caption{(Color online) Band structures of graphene bilayer nanoribbons with stacking (a) AB;  and (b) AA. The width of the ribbons is equal to 8 dimer chains in both cases.
}
\label{FigABAA}
\end{figure}

In Figs. \ref{FigABAA} (a) and \ref{FigABAA} (b) we show the bands of zigzag ribbons with AB and AA stacking respectively, calculated with the same model used for the twisted bilayer nanoribbon (TBNR).
By simply counting the edges in each system, four edge bands are expected for both cases \cite{Jaskolski_2011}. 
Indeed, for AB stacking 
there are four edge bands clearly identified in Fig. \ref{FigABAA} (a), which are degenerate at the first Brillouin zone boundary at zero energy. The bands for the AA-stacked ribbons show two Dirac points shifted from the BZ boundary; in this case, the edge bands are flat and split by $\pm \gamma_1$, if only vertical hoppings are considered. Within the model employed in this work, which
takes into account interlayer hoppings within a circle of radius $6a_{CC}$,
the splitting gives then an idea of the effective interlayer coupling, 
 that it is related to the bonding-antibonding nature of the AA states.

 Thus, in AB-stacked edges a large LDOS at $E_F$ (0 eV)  can be expected, while for edges with AA stacking, the peaks in the LDOS due to the edges are expected to be shifted below $E_F$ by an energy roughly given by the interlayer coupling.

 These energy differences explains why in the STM scans at low bias the bright spots at the edges, the so-called beads, are
 shifted with respect to the bright areas of the bulk moir\'e pattern 
(AA-stacked regions).  The bulk regions of the moir\'e pattern are bright in  STM images due to the larger LDOS near $E_F$. However, close to the edges,
 the maximum in the LDOS shifts to the AB sites. The low-bias topographic images shown in Fig. \ref{Fig1} are consistent with this result, as well as with the calculated LDOS
 of the ribbon
shown in Fig. \ref{Figribbon}.  Since the STM images were collected with a bias voltage
 of  100 meV, the  bright beads appear only in the AB edges, which are the sites of the only accessible states at
 that tip bias.
\begin{figure}[htbp]
\includegraphics[width=\columnwidth,clip]{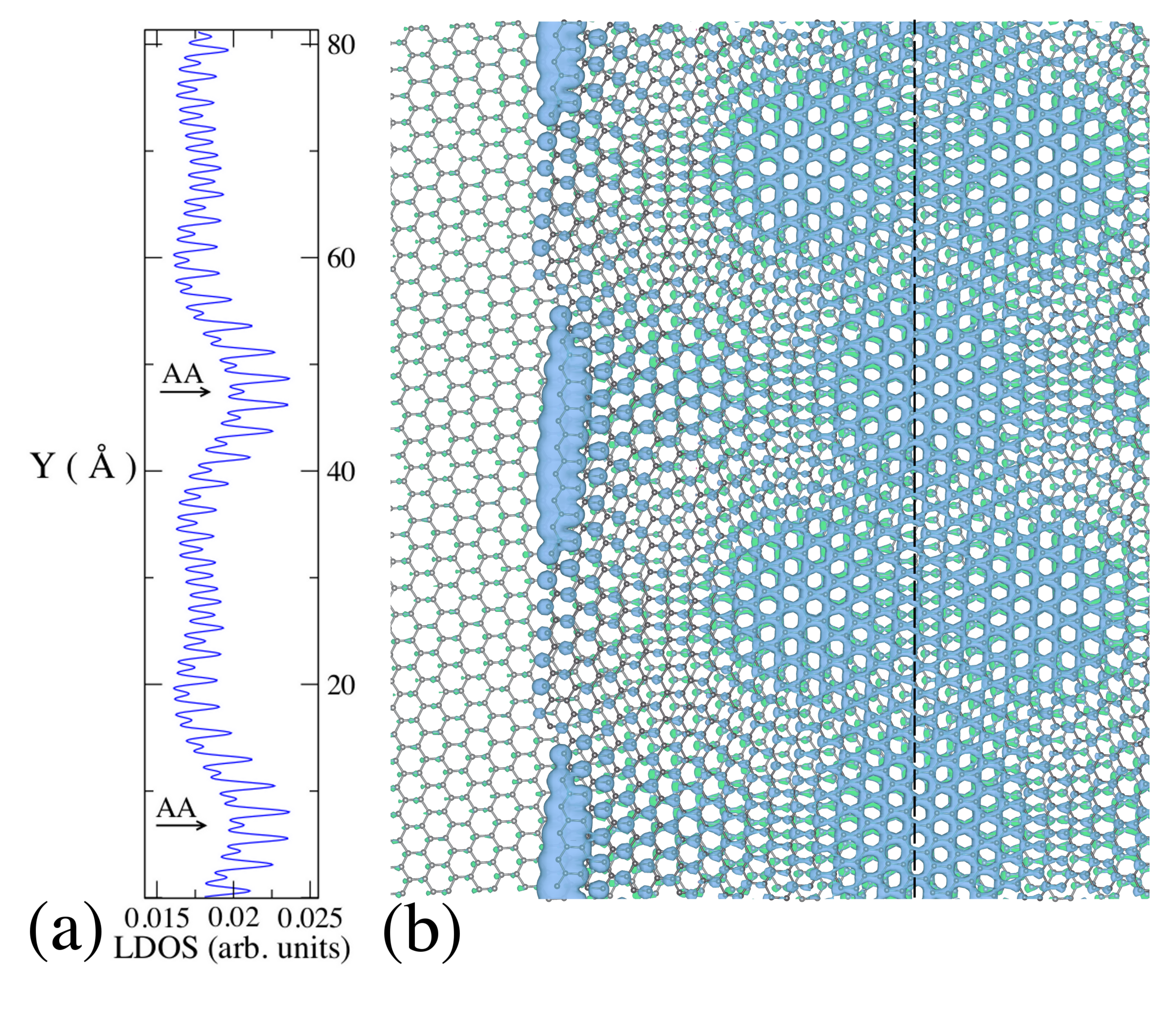}
\caption{(Color online)  STM (LDOS) simulation of a wide flake over graphene with energies E in the ranges $E_{F}-500$ meV < E < $E_{F}$.  (a) STM line profile along the center of the flake. (b) The dashed line show the direction of the line profile in (a). Coexistence of edge states and bulk moire pattern can be observed. }
\label{FigMoire}
\end{figure}

Our model also describes adequately
the coexistence of edge and bulk moir\'e states.
To display this feature we have 
calculated the electronic structure of
a very wide graphene flake on top of graphite,
in order to obtain also the
 bulk states. 
Fig. \ref{FigMoire} shows the LDOS for such a system
integrated in
an energy range between $E_F -500$ meV and $E_F$. The line profile in Fig. \ref{FigMoire} (a) of the STM simulation
is taken along the center of the flake. It shows that states at the AA sites of the bulk are
the brightest.
Here the arrows 
indicate the position of
AA sites of the bulk. In
Fig. \ref{FigMoire} (b) the coexistence of both types of states can be observed. The edge states are observed to be misaligned with respect to the moir\'e pattern displayed by the rotated layer, 
as it also 
happens 
to 
the experimental results shown in Fig. \ref{Fig1}.

\section{Conclusions}
\label{sec:sum}
In
this work 
we have addressed a phenomenon 
observed for several years,  but
not properly explained to date.  By using a TB model to describe the chiral edge states in moir\'e nanoribbons,
 we show that bead-like edge states correspond to AB regions of the moir\'e edges in graphite superstructures. The electrons with low energies in the bulk are localized in  AA regions, while at the edges
they remain confined at
  AB-stacked regions.
This explanation considers the usual polarization voltages used in STM measurements, 
i.e.,  low values around 100 mV.  
For higher bias voltages the calculation predicts less pronounced differences between AA and AB regions in 
STM images.  Consistently, a mixture of both types of states has also been
  observed experimentally.

\begin{acknowledgments}
This work has been partially supported by Spanish Ministry of Economy and Competitiveness under grant FIS2012-33521, and Chilean FONDECYT grants 1130950 and 1110935.

\end{acknowledgments}

%

\end{document}